\documentclass[preprint,showpacs,superscriptaddress,10pt]{revtex4-1} 
\usepackage{amsmath,amssymb,graphicx}
\begin{document}

\title{Single-pixel ghost microscopy based on compressed sensing and complementary modulation}
\author{Wen-Kai Yu}
\affiliation{Key Laboratory of Electronics and Information Technology for Space System, Center for Space Science and Applied Research, Chinese Academy of Sciences, Beijing 100190, China}%
\affiliation{University of Chinese Academy of Sciences, Beijing 100049, China}

\author{Xu-Ri Yao}
\affiliation{Key Laboratory of Electronics and Information Technology for Space System, Center for Space Science and Applied Research, Chinese Academy of Sciences, Beijing 100190, China}%
\affiliation{University of Chinese Academy of Sciences, Beijing 100049, China}

\author{Xue-Feng Liu}
\affiliation{Key Laboratory of Electronics and Information Technology for Space System, Center for Space Science and Applied Research, Chinese Academy of Sciences, Beijing 100190, China}%

\author{Long-Zhen Li}
\affiliation{Key Laboratory of Electronics and Information Technology for Space System, Center for Space Science and Applied Research, Chinese Academy of Sciences, Beijing 100190, China}%
\affiliation{University of Chinese Academy of Sciences, Beijing 100049, China}

\author{Ling-An Wu}
\affiliation{Laboratory of Optical Physics, Institute of Physics and Beijing National Laboratory for Condensed Matter Physics, Chinese Academy of Sciences, Beijing 100190, China}%

\author{Guang-Jie Zhai}
 \email{gjzhai@nssc.ac.cn}
\affiliation{Key Laboratory of Electronics and Information Technology for Space System, Center for Space Science and Applied Research, Chinese Academy of Sciences, Beijing 100190, China}%



\begin{abstract}
An experiment demonstrating single-pixel single-arm complementary compressive microscopic ghost imaging based on a digital micromirror device (DMD) has been performed. To solve the difficulty of projecting speckles or modulated light patterns onto tiny biological objects, we instead focus the microscopic image onto the DMD. With this system, we have successfully obtained a magnified image of micron-sized objects illuminated by the microscope's own incandescent lamp. The image quality of our scheme is more than an order of magnitude better than that obtained by conventional compressed sensing with the same total sampling rate, and moreover, the system is robust against intensity instabilities of the light source and may be used under very weak light conditions. Since only one reflection direction of the DMD is used, the other reflection arm is left open for future infrared light sampling. This represents a big step forward toward the practical application of compressive microscopic ghost imaging in the biological and material science fields.
\end{abstract}

\maketitle 
\section{Introduction}

Ghost imaging (GI) emerged two decades ago as a technique based on the correlated nature of both classical and quantum fields. Initially, ghost images were formed from two correlated light fields and two photodetectors: a high spatial-resolution detector used to record one light field which had not interacted with an object to be imaged, and a bucket detector without spatial resolution which was used to collect the light coming from the object. Neither detector can produce an image on its own, but a ``ghost" image can be constructed by correlating the signals of these two detectors. The first demonstration of ghost imaging (GI) \cite{Pittman1995,Strekalov1995} utilized biphoton pairs which were produced by spontaneous parametric down conversion in a nonlinear crystal, hence the phenomenon was interpreted as the result of quantum entanglement of the photon pairs \cite{Gatti2003}. Subsequently, further theoretical and experimental work showed that GI is also achievable with pseudo-thermal \cite{Boyd2002,Lugiato2004,Shih2005,Lugiato2005,Shih2006,Lugiato2006,Shih2008} as well as true thermal light \cite{Zhang2005,Liu2014} and can be explained with a classical statistical model \cite{Ferri2008,Zerom2012}.

Later, it was realized that the purpose of the reference beam was merely to measure the field distribution at the object, so it could just as well be replaced by a single predetermined spatially modulated object beam plus the bucket (single-pixel) detector. This modified technique was called computational GI by Shapiro \cite{Shapiro2008} and demonstrated by Bromberg et al. \cite{Bromberg2009}. But a few months earlier, the same optical protocol had been proposed and demonstrated independently by Baraniuk et al., who combined it with the compressive sampling (compressed sensing) algorithm \cite{Romberg2006,Donoho2006,Candes2006}, and called it ``Single-pixel imaging via compressive sampling" \cite{Baraniuk2008,Mittleman2008}. The only difference between single-pixel compressed sensing (CS) imaging and computational GI was that the spatial modulator was placed behind the object instead of in front.

It is known that traditional GI reconstruction algorithms are linear while CS algorithms are nonlinear. In CS, optimization is applied to recover a signal from incomplete or noisy observations of the original signal through random projections, provided that the signal is sparse or compressive in some basis. The application of CS in the field of imaging allows one to retrieve high resolution images from many fewer measurements than those established by the Nyquist-Shannon criterion. Recently, the GI community has employed CS to obtain compressed ghost images \cite{Katz2009}, greatly reducing both the acquisition time and the number of frames required. These improvements have motivated an ongoing effort to implement technologies based on GI such as multispectral imaging \cite{Lu2011}, optical encryption \cite{ShenLi2013,Yu2013}, and most recently, adaptive compressive GI \cite{YuOE2014}. Although GI and single-pixel compressive imaging were historically independent and developed in parallel with each other, both techniques promise a resource-efficient alternative to array detectors, permitting us to reduce operational problems involved in systems based on raster scanning. However, practical random patterns sequentially fed into a DMD only have two values, 0 or 1, corresponding to the $\pm 12 ^\circ$ angles at which the micromirrors are deflected, which is not the best range of values for compressive GI. It is interesting to note that when both reflected beams from the DMD are used to reconstruct a ghost image, the corresponding frames are complementary. Accordingly, instead of wasting the 0 ($-12 ^\circ$ reflected) beam we have developed a novel technique which we call complementary compressive imaging, that makes full use of both reflections to dramatically improve the image quality \cite{YuSR2014}.

On another note, some time ago Luo et al. \cite{Luo2011,Luo2012} reported a technique that they called correspondence imaging, in which a positive or negative image is retrieved by conditional averaging of the reference signals; that is, only those reference data that correspond to positive or negative intensity fluctuations of the bucket signal are selected for simple averaging, without the need to multiply by the bucket detector intensity itself. It seemed to defy intuition in that no direct second-order correlation calculation is performed, while compared with conventional GI the number of exposures used to reconstruct the images and consequently the computation time are greatly reduced. Further developments of this positive-negative image concept have been presented in many other papers \cite{Lipra2013,Liapl2013,Shih2012,Sun1,Sun2}. By alternately measuring the bucket values of a random binary pattern and its inverse, Sun et al. \cite{Sun1,Sun2} reconstructed a ghost image by correlating non-inverted patterns with the differential signals of the complementary illuminated pairs, and normalized the bucket signal with respect to the positive or negative intensity fluctuations averaged to 0. Their method was essentially the same as that of correspondence imaging, but their GI bucket values were used as a series of weighting factors instead of just 0 and 1. Their projected inverse patterns were only used in generating the differential bucket values and not in the correlation computation, so their scheme is not suitable for compressive GI, which is a nonlinear approach and needs a one-to-one relationship between the modulation patterns and measured values. Moreover, it is very difficult to project black and white speckles onto tiny biological tissues by a projector. To our knowledge, the only previous report on the use of GI in actual microscopy was based on entangled photon pairs, which is a complicated and expensive light source \cite{Nasr2009}. Ghost imaging with a classical thermal source was studied in a two-arm microscope imaging system but it was only a theoretical simulation with a simple double-slit object \cite{Bai2012}. Recently, CS was also applied to microscopy \cite{YuehaoWu2010,Studer2012}; in \cite{Studer2012}, Studer et al. tested their system on a sample of fluorescent beads which were sparsely distributed. In addition, they used binary patterns of a shifted and rescaled form $(1+h)/2$ where $h$ is a Hadamard sequence, thus each entry of patterns is either 0 or 1, still not the best range of values. In \cite{Radwell2014}, Radwell et al. proposed an ``adaptive" mode that provides the benefits of both high frame rate and high resolution, but still used Hadamard patterns for image reconstruction of a simple silicon CMOS chip. Here, as an alternative to complementary measurements in both reflection arms as in \cite{YuSR2014}, a prototype single-pixel microscope with complementary modulation is presented. It can not only image complex gray samples but also produce very satisfactory image quality, more than an order of magnitude better than traditional compressive imaging. Moreover, since only one reflection direction of the DMD is used, the other reflection arm is left open for future infrared light simultaneous sampling (which has just been demonstrated by Radwell et al. \cite{Radwell2014}).

On the basis of our former work \cite{Luo2011,Luo2012,YuSR2014}, we have extended our complementary compressive imaging technique to the imaging of real biological samples using an ordinary microscope's own illumination lamp. In our proof-of-principle experiment, the object's image is projected onto a DMD, which is encoded with randomly modulated array patterns alternated with their inverse (complementary) patterns. Moreover, we use only one reflection arm with a single-pixel ultra-weak light detector rather than the two detectors in \cite{YuSR2014}, leaving room for future simultaneous infrared light detection in the other arm. Through our method, complex gray-scale images of the mouthparts of a female mosquito have been reconstructed with far fewer measurements and a performance more than an order of magnitude better than that of traditional CS, opening exciting prospects for the important field of microscopic imaging.

\section{Experimental Setup and Implementation}

\begin{figure}[htbp]
\centering\fbox{\includegraphics[width=7cm]{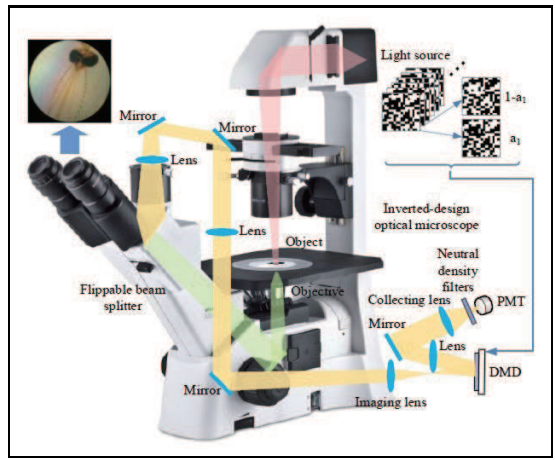}}
\caption{Schematic of the experimental setup.}
\label{fig:setup}
\end{figure}

The experimental apparatus is illustrated in Fig.~\ref{fig:setup}. The thermal light from a halogen lamp illuminates the sample table from overhead with an ordinary incandescent lamp (a noncoherent thermal source). Light from the object passes vertically down through the objective, and after reaching a flippable beam splitter may be viewed directly through the eyepiece or transmitted by various mirrors and lenses to the DMD. A $10\times$ magnification is achievable with just the objective, which has a numerical aperture of 0.25. The DMD consists of $768\times 1,024$ micro-mirrors each of size $13.68\times13.68\ \mu\textrm{m}^2$. Each mirror rotates about a hinge and can be switched between two positions oriented at $+12 ^{\circ}$ or $-12 ^{\circ}$ with respect to the plane of the DMD; the $+12 ^{\circ}$ position corresponds to bright pixels 1, and inversely, $-12 ^{\circ}$ to dark pixels 0. The modulation frequency of the DMD can reach 32.5 kHz. A photomultiplier tube (PMT) (Hamamatsu H7468-20) is used as the bucket (single-pixel) detector for collecting the total light intensity reflected from the DMD, with some neutral density filters in front for suitable attenuation. Therefore, our proof-of-principle single-pixel microscopic imaging experiment is performed under ultra-weak light conditions where ordinary microscopy photography would be impossible. Generally, in biological microscopy, the object is a fluorescing sample like cells, tissues and beads, the light from which is extremely weak. Our sample consisted of the mouthparts of a female mosquito, of about $0.8\times 2.4\ \textrm{mm}^2$ in size. In order to show what the object is and to make sure the image is clear, we first took a photograph of the mosquito sample (as illustrated in the upper left corner of Fig.~\ref{fig:setup}) as a reference image directly in front of the eyepiece of the microscope with an ordinary charge coupled device (CCD) monitor and a common flippable beam splitter, under bright illumination.

In our configuration the size of the image to be reconstructed is $768\times 1,024$ pixels, equal to that of the DMD. However, the time consumed by CS increases exponentially with the dimensions, so to recover such a large image we performed a row-scanning strategy: the DMD was activated row by row and the same number of complementary exposures were taken for each row. Since the light source can be considered to be stable within two reasonably short modulation time intervals, we modulated the complementary frame pairs of each row alternately. This was repeated until all rows had been measured. Then we calculated the differential values of adjacent elements (in a one-to-one complementary relationship) of the bucket signal with odd and even index numbers, corresponding to complementary differential frames. After all rows had been reconstructed, the whole image was retrieved via the image mosaic method.

\section{Complementary Compressive Ghost Imaging Framework and Results}

In conventional GI, an image of pixel size $p\times q$ can be acquired by correlating the reference frame intensities $I_R$ with the bucket detector signal $S_B$:
\begin{eqnarray}
\textrm{GI}=\langle S_B I_R(x_R)\rangle,
\label{eq:1}
\end{eqnarray}
where $\langle\cdot\cdot\cdot\rangle$ is the average operator, and $x_R$ denotes the spatial position in the reference detector frames. Each bucket detector value can be defined as
\begin{eqnarray}
S_B=\int_{\Omega}I_R(x_i,x_j)T(x_i,x_j)dx_idx_j,
\label{eq:2}
\end{eqnarray}
where $x_i$, $x_j$ denote the row and column indices, respectively, $T$ represents the intensity transmission function of the object, and $\Omega$ is the area of the beam.

In compressive GI, we can flatten the two-dimensional image into a one-dimensional column vector, and denote it by $x$ of length $N$, where $N=p\times q$. Each pattern of the DMD can also be reshaped into a row vector, and then $M$ such vectors re-arranged into a measurement matrix $A$ of size $M\times N$. Now the sequence of bucket values is also given as a column vector $y$ of size $M\times 1$, thus the problem can be described by the set of linear equations:
\begin{eqnarray}
y=Ax+e,
\label{eq:3}
\end{eqnarray}
where $e$ of size $M\times 1$ denotes the noise. When $M\geq N$, these equations can be easily solved by various iterative methods, such as lower upper decomposition, Gauss-Jordan elimination, Cholesky decomposition, or singular value decomposition. However, as the image size increases, the computational overhead grows rapidly. It is known that a natural image generally has a sparse representation in a certain basis $\Psi$ (e.g. Haar wavelet basis, discrete cosine transform basis, or Fourier transform basis). This prior knowledge is very helpful for addressing an ill-conditioned problem where the number of measurements is fewer than the total number of pixels $N$, and for reducing the number of measurements required for the image acquisition. Here we expand $x$ in an orthogonal basis $\Psi=[\psi_1,\psi_2,...,\psi_N]$ as follows:
\begin{eqnarray}
x=\Psi x',\ \textrm{or}\ x=\sum\limits_{i = 1}^N{{x'}_i}{\psi_i},
\label{eq:4}
\end{eqnarray}
where $x'$ of size $N\times 1$ is the coefficient sequence of $x$. We say that $x'$ is $k$-sparse if $\left|{\left\{{i:{x'}_i\ne 0,1\leq i\leq N}\right\}}\right| \le k$. Then the problem becomes
\begin{eqnarray}
y=A\Psi x'+e.
\label{eq:5}
\end{eqnarray}

For faithful image reconstruction, rows of $A$ should be incoherent with the sparse basis. It is found that a completely random matrix $A$ works well, from which accurate recovery is possible by convex optimization \cite{Candes2006}:
\begin{eqnarray}
\mathop{\min}\limits_{x'}\frac{1}{2}\left\|{y-A\Psi x'}\right\|_2^2+\tau {\left\|{x'}\right\|_1},
\label{eq:6}
\end{eqnarray}
where $\tau$ is a constant scalar weighting the relative
strength of the two terms, and $\left\|\cdot\cdot\cdot\right\|_p$ stands for $l_p$ norm defined as ${\left({\left\|x\right\|}_p\right)^p}= {\sum\nolimits_{i=1}^N{\left|{x_i}\right|}^p}$. The first term is small when the optimal $x'$ is consistent with Eq.~(\ref{eq:5}) within a small error. The second term is small when $x'$ is sparse. Recent research \cite{CBLi2010} has proved that the use of total variation (TV) regularization instead of the $l_1$ term in CS problems gives a sharper recovered image by preserving the edges or boundaries more accurately, and the gradient of an image is generally sparse as well. Therefore, we use the TVAL3 solver \cite{CBLi2010} here.

According to the method proposed by Sun et al. \cite{Sun1,Sun2}, ghost images can be retrieved by correlating differential signals $\Delta S_B$ and non-inverted patterns $I_R(x_R)$. That is, $\textrm{GI}_\textrm{C}=\langle\Delta S_BI_R(x_R)\rangle$. The differential bucket signal in the second-order correlation form can be viewed as weighting factors, which indirectly gives the correlation coefficient between $I_R(x_R)$ and the object (i.e., $\Delta S_B>0$ and $\Delta S_B<0$ indicate a positive and a negative linear relationship, respectively, and $\Delta S_B=0$ means no correlation), thus their method has a performance somewhat better than traditional second-order correlation but actually much poorer than that of compressive GI with the same number of measurements. Although the problem to be solved whether using intensity correlation or CS nonlinear algorithms is the same, the former correlation fails to work in CS because it destroys a precise correspondence relationship.

We now consider the complementary sampling scheme proposed in our previous work \cite{YuSR2014}. Since each complementary frame pair appears alternately, the odd and the even subscripts constitute two supporting mutually related sets: $odd=\{1,3,5,\ldots\}$, and $even=\{2,4,6,\ldots\}$. We denote the measured bucket signal and the measurement matrix used as $y'$ and $A'$, respectively, then the complementary differential bucket signal is defined as
\begin{eqnarray}
y=\frac{1}{2}(y'_{odd}-y'_{even}).
\label{eq:7}
\end{eqnarray}
Here, $y$ decreases the variance of the independent and identically distributed noise by a half \cite{YuSR2014}. Accordingly, the complementary differential frames can be written as
\begin{eqnarray}
A=\frac{1}{2}(A'_{odd}-A'_{even})=A'_{odd}-\frac{1}{2}_{M\times N},
\label{eq:8}
\end{eqnarray}
where the inverse frames $A'_{even}=1_{M\times N}-A'_{odd}$, $\frac{1}{2}_{M\times N}$ stand for an array of all $\frac{1}{2}$, and we have let $A'_{odd}$ and $A'_{even}$ denote the submatrix of $A'$ that contains all the columns corresponding to the support sets $odd$ and $even$. Notice that $A_o$ and $A_e$ on the DMD are binary matrices consisting of the two values 0 or 1, and thus $A$ is a $\pm0.5$ binary matrix, which generates the ``positive-negative" light modulation. Then Eq.~(\ref{eq:5}) can be rewritten as
\begin{eqnarray}
\frac{1}{2}(y'_{odd}-y'_{even})=(A'_{odd}-\frac{1}{2}_{M\times N})\Psi x'+\frac{1}{2}(e_1-e_2).
\label{eq:9}
\end{eqnarray}

\begin{figure}[htbp]
\centering\fbox{\includegraphics[width=5.5cm]{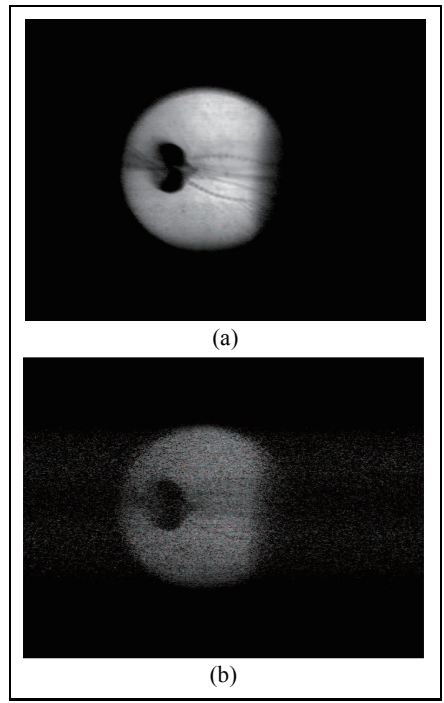}}
\caption{Reconstructed images of the mosquito mouthparts of pixel size $768\times1,024$. Image retrieved by (a) complementary CS, and (b) conventional CS.}
\label{fig:Expresult}
\end{figure}

By utilising this technique, we have reconstructed the image of the mouthparts of a female mosquito, as pictured in Fig.~\ref{fig:Expresult}(a), in which even the flagellum, clavola, upper and lower jaws, and labium of the mouthparts can be seen clearly. For comparison, we also show the image recovered via traditional CS (see Fig.~\ref{fig:Expresult}(b)) with the same total sampling rate of 50 \% (that is, complementary CS samples 256 pattern pairs (i.e., 512 patterns in total) for each row of length 1024, compared to 512 patterns for conventional CS). From the experimental results, we can see that the image quality of our new method is much better than that of conventional CS.

\section{Performance and discussion}

\begin{figure}[htbp]
\centering\fbox{\includegraphics[width=6.5cm]{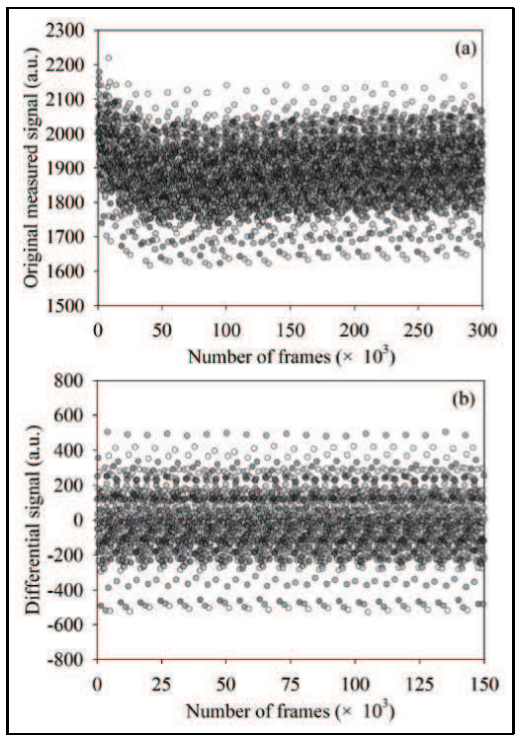}}
\caption{Simulation of the effect of intensity instabilities on the bucket detector fluctuations: (a) $y'$ and (b) $y$.}
\label{fig:analysis}
\end{figure}

In our earlier work \cite{YuSR2014}, we already proved by simulation that the total number of measurements required in the two-detector complementary sampling scheme is much smaller than that in traditional CS methods for the same mean square error (MSE) value, although the former needs to measure the signals in both reflection arms of the DMD. Moreover, it was also demonstrated that the computation time is shorter than that of traditional CS. As the system proposed in this paper is an extension of our previous scheme to the microscopic regime but with only one detector arm, it embodies all the same advantages. Since the modulation frequency used here is 450~Hz and the total number of measurements is the same, the total acquisition time for both complementary CS and conventional CS is altogether 14.56 minutes. However, for the same image quality, conventional CS will require a significant increase in the number of exposures and consequently measurement time. Furthermore, an added bonus is the robustness of complementary CS against any instability of the light output from the lamp of an actual microscopic imaging system, which would affect the bucket detector signal $y'$. To simulate the deleterious effect of unstable illumination, we programmed the DMD such that the central row (Row 384) displayed 616 random patterns repetitively, as a result of which the measured bucket detector signal showed corresponding periodic variations (see Fig.~\ref{fig:analysis}(a)). This instability would result in poor image quality with conventional CS. However, by the pairwise subtraction algorithm within each complementary pair, the bucket detector values are converted into a ``positive-negative" intensity distribution vector with a mean of $\sim$ 0, as shown in Fig.~\ref{fig:analysis}(b), thus compensating for any undesirable fluctuations of the average illumination intensity.

To obtain a more quantitative measure of the reconstructed image quality, the peak signal-to-noise ratio (PSNR) is defined here as a figure of merit:
\begin{eqnarray}
\textrm{PSNR}=10\log(255^2/\textrm{MSE}),
\label{eq:10}
\end{eqnarray}
where
\begin{eqnarray}
\textrm{MSE}=\frac{1}{pq}\sum\nolimits_{r,c=1}^{p,q}[U_o(r,c)-\tilde U(r,c)]^2.
\label{eq:11}
\end{eqnarray}
This MSE describes the distance squared between the reconstructed image $\tilde U$ and the original image $U_o$ for all $p\times q$ pixels. Naturally, the larger the PSNR value is, the better the quality of the image recovered.

\begin{figure}[htbp]
\centering\fbox{\includegraphics[width=6.5cm]{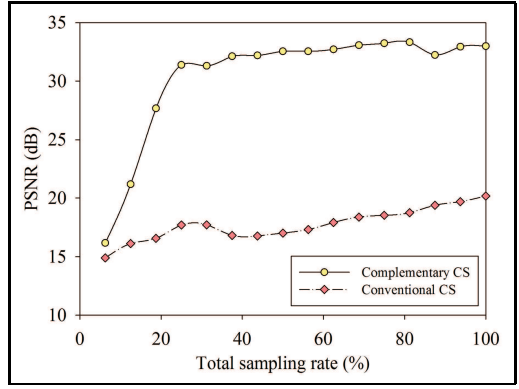}}
\caption{PSNR values of complementary CS and conventional CS versus the total sampling rate.}
\label{fig:PSNRcurve}
\end{figure}

In Fig.~\ref{fig:PSNRcurve} the PSNR values of complementary CS and conventional CS are plotted as a function of the total sampling rate, which is defined as the ratio between the total number of measurements and the length of the image signal. As shown in the figure, (yellow circles: complementary CS, pink diamonds: conventional CS), for both methods the image quality increases with the sampling rate, tending to saturation at a high level. It is clearly visible that the image quality of complementary CS is more than an order of magnitude better than that of conventional CS with the same total sampling rate, i.e., the number of measurements is significantly reduced for the same image quality. The maximum difference in the PSNR is more than 15 dB, i.e., corresponding to a maximum MSE difference of about 32.

Since the values of $A_o$ and $A_e$ all consist of either 0 or 1, and obey the Bernoulli distribution, the mean of both matrices is always close to 0.5. The impact of the background signal $y_b=\frac{1}{2}_{M\times N}\Psi x'$ (which can also be viewed as the background noise) on image quality should not be neglected. Indeed, our Eq.~(\ref{eq:9}) subtracts this background term and gives a better performance than Eq.~(\ref{eq:5}). Moreover, the mean of $A$ is close to 0, which in fact helps to pick out useful information from the bucket signal.

\begin{figure}[htbp]
\centering\fbox{\includegraphics[width=6cm]{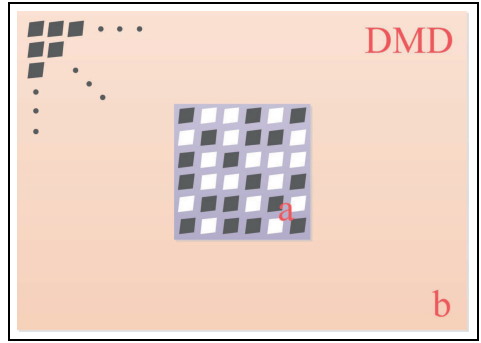}}
\caption{Schematic diagram of the DMD. The micromirrors in the $+12 ^{\circ}$ ($-12^{\circ}$) orientation are shown in white (black). The lilac and peach colored areas represent the modulated and the nonactive regions of the DMD, respectively.}
\label{fig:DMD}
\end{figure}

Another source of noise which we should consider is the DMD. As illustrated in Fig.~\ref{fig:DMD}, the random patterns are only encoded onto the activated area $a$ of the DMD, while all the other micromirrors in the remaining region $b$ are in the $-12 ^\circ$ position. The gaps between adjacent micromirrors and gaps incurred during flipping of the micromirrors which expose the metal surface beneath will incur some random noise, including mirror-like and diffuse reflections. The former will not be reflected into the direction of detection and so can be neglected. However, the noise from dark pixels in area $b$ and from gaps generated by micromirror flipping within the image area is worthy of attention. To quantify this noise, we need to add the following extra eight measurements:
\begin{eqnarray}
\left\{\begin{array}{l}
{a_+}+{b_-}={f_1}\\
{a_+}+{b_+}={f_2}\\
{a_-}+{b_-}={f_3}\\
{a_-}+{b_+}={f_4}\\
a_+^{object}+b_-^{object}={f_5}\\
a_+^{object}+b_+^{object}={f_6}\\
a_-^{object}+b_-^{object}={f_7}\\
a_-^{object}+b_+^{object}={f_8},
\end{array}\right.
\label{eq:12}
\end{eqnarray}
where $a$, $b$ denote the signals from modulated and the nonactive regions of the DMD, and the subscripts ``$+$" and ``$-$" denote the states when all micromirrors are aligned in the $+12 ^\circ$ or $-12 ^\circ$ positions, respectively. The superscript ``$object$" means that the object lies within the detector beam path, otherwise it is outside. Here $f_1, f_2, \ldots, f_8$ are the corresponding measured values of the eight equations. By solving these equations, we can easily calculate $b_-^{object}$, $a_+^{object}$, $a_+$ and $a_-^{object}$. Considering the dark noise $c_{dark}$ of the PMT, we can rewrite Eq.~(\ref{eq:5}) as
\begin{eqnarray}
y=A\Psi x'+(c_{dark}+b_-^{object}+0.5\frac{a_+^{object}}{a_+}a_-^{object}),
\label{eq:13}
\end{eqnarray}
where $t=\frac{a_+^{object}}{a_+}$ denotes the total intensity transmission of the object. Sinse the DMD pixels are randomly modulated, only half of the light flux in area $a$ is detected in each exposure, so we only need discuss the noise in half of the (object's) transparent part. Except for $c_{dark}$ which is a variable, the other two terms ($b_-^{object}$ and $0.5\frac{a_+^{object}}{a_+}a_-^{object}$) in the noise $e$ are all constants, and thus can be subtracted. However, even after these two items are subtracted, the background noise $\frac{1}{2}_{M\times N}\Psi x'$ mentioned above still plays a primary role in image degradation and the image quality is not expected to be improved dramatically. Since the dark noise is independently and identically distributed in both $y'_{odd}$ and $y'_{even}$, complementary CS can not only decrease the variance of the dark noise by half but also remove the impact of the other two kinds of noise ($b_-^{object}$ and $0.5\frac{a_+^{object}}{a_+}a_-^{object}$), via Eq.~(\ref{eq:9}). Although the complementary differential bucket signals seem to be a subtracted and rescaled version of those in conventional CS imaging, our method actually averages the variance of the noise, thus significantly improving the SNR in the measurement process.

\section{Conclusions and Prospects}

In summary, we have proposed and experimentally demonstrated a single-pixel single-arm DMD-based compressive ghost microscopy system, based on direct complementary modulation of the DMD. In order to incorporate the microscope into this system, the image is directly focused onto the DMD plane, rather than using structured illumination by a digital projector or two-arm sampling of traditional GI. The mouthparts of a female mosquito have been successfully imaged and identified. An image quality more than an order of magnitude better than that obtained by conventional CS with the same total sampling rate has been realized, and moreover, the system is very robust against intensity instabilities of the illumination source, while imaging is still achievable under very weak light conditions. Compared with the $\pm12\ ^{\circ}$ double-arm detection scheme in our previous work \cite{YuSR2014}, it allows room for future simultaneous near-infrared light detection in the other reflection arm. Our scheme should open exciting prospects for microscopy in the biosciences and material sciences, notably for biological imaging of sparse fluorescent objects such as cells and tissues.

\textbf{Note:}

Here it may be interesting to recall the submission history of our paper. We first submitted our manuscript to Appl. Phys. Lett. (Doc. ID: L14-04737) on 25 May 2014. The editors told us that it did not meet the timeliness requirement for rapid publication in their journal, so we submitted the manuscript to Opt. Express (Doc. ID: 213211) on 31 May 2014. We received the first review decision on 30 June 2014, which was rejection on the grounds that the earliest idea of using a DMD instead of an SLM in computational ghost imaging was reported in the paper by Sun et al. \cite{Sun1}, and the same method of using opposite light pattern pairs in another conference paper by Sun et al. \cite{Sun2}, where they acquired the differential signals of each illumination pair for correlation calculation, rather than absolute signals. Actually, the use of a DMD instead of an SLM in computational GI is nothing new and was reported in many papers long before Sun's, e.g. \cite{Lu2011,ShenLi2013,Yu2013}. In addition, as mentioned in Introduction, Sun's method was essentially the same as that of ``correspondence imaging", first proposed by Luo et al. in \cite{Luo2011,Luo2012}, but their GI bucket values were used as a series of weighting factors instead of just 0 and 1.

We then modified the content of our manuscript and submitted it to Optics Letters (Doc. ID: 216782) on 10 July 2014. We received the first manuscript decision on 25 July 2014. This time, a referee recommended that we should explain more clearly how and why the complementary pattern strategy is better than prior work on the linear reconstruction of images like in traditional GI, or the nonlinear approach of single-pixel imaging.

On this advice, we added substantially more content to our manuscript, which then greatly exceeded the length restrictions of Optics Letters, so we once again submitted it to Optics Express (Doc. ID: 224607) on 7 Oct 2014. We received the first decision on 28 Oct 2014, but strongly disagreed with the reviewer's comments and appealed for another review, which was granted on 5 Nov 2014. We received the review decision on 7 Jan 2015. The new reviewer suggest that the manuscript can be published only after we clear up the picture and the presentation of ``ghost imaging", and present it as the compressive imaging aspect only. However, we think that the only difference between single-pixel compressive imaging \cite{Baraniuk2008} and computational GI \cite{Shapiro2008} was that the spatial modulator was placed behind the object instead of in front. Furthermore, the GI community has already employed CS to obtain compressed ghost images (see in \cite{Katz2009,Lu2011,Yu2013,YuOE2014}, Opt. Lett. \textbf{37}, 1067--1069, 2012, Phys. Lett. A \textbf{376}, 1519--1522, 2012, Phys. Lett. A, \textbf{377}, 1844--1847, 2013). Most intriguingly, the CS community also developed single-pixel compressive architecture based on an active illumination concept, as demonstrated in \cite{YuehaoWu2010} and Appl. Opt. \textbf{405}, 405--414, 2011, where the spatial modulator was placed in front of the object, just as the same with the optical layout of computational GI. Removing the compressive aspect, i.e. taking the measurement matrix to be full rank, this imaging scheme turns into a mechanism named multiplexed illumination (in: Proc. IEEE International Conference on Computer Vision, 2003, 808–-815). For example, in Hadamard-multiplexed illumination, about half of the light sources operate simultaneously, creating brighter, clearer captured images. In fact, the mathematical models of single-pixel compressive imaging, computational GI or multiplexed illumination are the same \cite{Katz2009}, only with the change of reconstruction algorithms, which will in turn result in the difference of the number of measurements. Although GI, single-pixel compressive imaging and multiplexed illumination were historically independent and developed in parallel with each other, these techniques promise a resource-efficient alternative to array detectors, permitting us to reduce operational problems involved in systems based on raster scanning.

Besides, Padgett's group should be congratulated on quickly realizing the dual infrared and visible microscope system, as described in \cite{Radwell2014}.

\section*{Acknowledgments}
National Key Scientific Instrument and Equipment Development Project of China (2013YQ030595); National High Technology Research and Development Program of China (2013AA122902); National Basic Research Program of China (2010CB922904); National Natural Science Foundation of China (11275024).


\end{document}